\documentclass[manuscript]{aastex}

\usepackage{graphicx}
\usepackage{rotating}
\usepackage{dcolumn}
\usepackage{natbib}

\newcommand{\rsun}{R_{\Sun}}
\newcommand{\mydeg}{^{\circ}}
\newcommand{\Rss}{R_{\mathrm{SS}}}
\shorttitle{CME Deflection and Angular Momentum}
\shortauthors{Kay \& Opher}

\begin{document}

\title{The Heliocentric Distance Where the Deflections and Rotations of Solar Coronal Mass Ejections Occur}

\author{C. Kay}
\affil{Astronomy Department, Boston University, Boston, MA 02215}
\email{ckay@bu.edu}

\and 

\author{M. Opher}
\affil{Astronomy Department, Boston University, Boston, MA 02215}

\begin{abstract}
Understanding the trajectory of a coronal mass ejection (CME), including any deflection from a radial path, and the orientation of its magnetic field is essential for space weather predictions.  \citet{Kay15} developed a model, Forecasting a CME's Altered Trajectory (ForeCAT), of CME deflections and rotation due to magnetic forces, not including the effects of reconnection.  ForeCAT is able to reproduce the deflection of observed CMEs \citep{Kay15L}.  The deflecting CMEs tend to show a rapid increase of their angular momentum close to the Sun, followed by little to no increase at farther distances.  Here we quantify the distance at which the CME deflection is ``determined,'' which we define as the distance after which the background solar wind has negligible influence on the total deflection.  We consider a wide range in CME masses and radial speeds and determine that the deflection and rotation of these CMEs can be well-described by assuming they propagate with constant angular momentum beyond 10 $\rsun$.  The assumption of constant angular momentum beyond 10 $\rsun$ yields underestimates of the total deflection at 1 AU of only 1\% to 5\% and underestimates of the rotation of 10\%.  Since the deflection from magnetic forces is determined by 10 $\rsun$, non-magnetic forces must be responsible for any observed interplanetary deflections or rotations where the CME has increasing angular momentum.\end{abstract}


\section{Introduction}
Knowing whether a coronal mass ejection (CME) will impact and the orientation of its magnetic field upon impact is critical for predicting space weather effects at Earth and throughout the heliosphere.  The intensity of CME-driven geomagnetic storms (as measured by $Dst$) increases with the magnitude of the CME velocity and the southward magnetic field strength \citep{Gop08, Xio06a, Xio07}.  Observations show that CMEs deflect, deviating from a purely radial trajectory, as well as rotate, changing their orientation.  Coronal observations show that extreme deflections can happen close to the Sun \citep{Mac86, Byr10, Gui11, Isa14}.  Several of the deflections presented in \citet{Isa14} exhibit latitudinal deflections exceeding 30$\mydeg$ below 8 $\rsun$.  The deflection motion can continue out into interplanetary space \citep{Wan04, Lug10a, Dav13, Isa14, Wan14}. Some of these interplanetary deflections result from interactions with CMEs or other obstacles \citep{Xio06b,Xio09,Lug12}, however many of the observed cases have no such association.  The exact relation between coronal and interplanetary deflections remains uncertain - the interplanetary CMEs may continue deflecting with constant angular momentum obtained in the corona or additional forces may cause acceleration at interplanetary distances.  

Early observations of CME deflections came from single coronagraph measurements and accordingly only latitudinal deflections were observed \citep{Hil77,Mac86, Cre04, Kil09}.  These observations showed that during solar minimum CMEs tend to deflect toward the solar equator, but the trend becomes less obvious during other times of the solar cycle.  More information on CMEs' longitudinal behavior became possible with the advent of the STEREO spacecrafts allowing for use of multiple coronagraph viewpoints and geometric reconstruction techniques.  These observations showed that deflections occur in longitude as well as latitude \citep{Liu10b, Lug10, Dav13, Isa13, Nie13, Isa14}.  \citet{Isa13} show that the latitudinal deflection tends to exceed the longitudinal deflection, however their CME sample only includes CMEs during solar minimum.  \citet{Wan14} show that, even for a solar minimum CME, interplanetary longitudinal deflections exceeding 30$\mydeg$ can occur.

Due to the large deflections in the low corona, where the magnetic pressure exceeds the thermal pressure, magnetic forces have become a popular explanation for the observed deflections.  \citet{Kil09} suggest that the open magnetic fields of coronal holes naturally guide CMEs down toward the equator.  \citet{She11} and \citet{Gui11} attribute the deflection to the gradients in the background solar magnetic energy.  Magnetic forces deflect a CME toward the region of lowest magnetic energy, which on global scales corresponds to streamer regions or, at farther distances, the Heliospheric Current Sheet (HCS).  The direction of the deflection then depends on the location of the HCS throughout the solar cycle.  During solar minimum, when the HCS is relatively flat, primarily latitudinal deflections should occur.  As the inclination of the HCS increases deflections will occur in a wider variety of directions.

\citet{Kay13} and \citet{Kay15} present a model, Forecasting a CME's Altered Trajectory (ForeCAT), for magnetic CME deflections.  Previous deflection models have simply compared the direction of the deflection and the magnetic gradients at distances greater than 2 $\rsun$.  ForeCAT simulates the deflection of CME from the integrated effect of the background forces beginning at the eruption of the CME.  ForeCAT includes both magnetic tension and magnetic pressure gradients, and non-radial drag that opposes the deflection motion, although \citet{Kay15} find that the non-radial drag has minor effects unless abnormally large drag coefficients are used. To describe the CME's radial propagation, ForeCAT uses a three-phase propagation model, similar to that present in \citet{Zha06}, which implicitly includes the effects of drag in the radial direction. The magnetic tension force tends to be of comparable magnitude and point in the same direction as the magnetic gradients.  ForeCAT reproduces both the global trends in the direction and magnitude of CME deflections \citep{Kay15} and the deflection of a specific observed case \citep{Kay15L}.

\citet{Kay15} found that the deflection forces quickly increase a CME's angular momentum as it begins deflecting, typically toward the HCS.  The magnetic forces decrease rapidly with distance causing the angular momentum to increase at a much smaller rate (see Figure 5 of \citet{Kay15}).  For a strong magnetic background (e.g. a CME erupting from a declining phase active region) the initial increase in the angular momentum greatly exceeds any additional angular momentum gained beyond a few solar radii.  For a weak magnetic background (e.g. a quiet Sun or solar minimum CME) the initial increase in the angular momentum is small and the slow continued increase at farther distances corresponds to a larger percentage of the total angular momentum gained by the CME by 1 AU. 

ForeCAT can also simulate CME rotation due to the differential forces acting upon the CME.  CME rotation is not as well observed as CME deflection but \citet{Vou11} infer an extreme rotation of nearly 80$\mydeg$ below 30 $\rsun$ for the 2010 June 16 CME.  In \citet{Kay15} we present ForeCAT's ability to determine rotation, but do not explore it in great detail.

Here we focus on the evolution of the angular momentum of CMEs erupting from strong magnetic backgrounds.  These CMEs tend to be the fastest and have the strongest magnetic field, leading to the more extreme space weather effects at Earth.  We expand the work of \citet{Kay15} with simulations of 200 new CMEs deflecting and rotating during propagation out to 1 AU.  We ascertain at what distance CME deflections and rotations are ``determined.''  We want to find the distance at which the deflection forces have a negligible influence on the CME's trajectory and orientation.  This corresponds to the distance beyond which the CME propagates with constant angular momentum.

\section{ForeCAT}
In the most recent version of ForeCAT \citep{Kay15} many of the simplifying assumptions were removed, most notably the restriction of the deflection to a single deflection plane.  ForeCAT calculates the magnetic forces due to the solar background across the surface of a torus representing the CME flux rope (see Figures 1 and 2 of \citet{Kay15}.  In ForeCAT, the deflection and the rotation are decoupled.  The net force creates a deflection corresponding to a translational motion of the center of mass of the CME.  This motion can be in any non-radial direction so the CME is free to deflect in three dimensions.  ForeCAT's one-dimensional rotation results from a net torque about the axis connecting the nose of the CME to the center of the Sun.  Rotation changes the tilt of the CME, but does not affect its latitude or longitude.  Observations show that CMEs can rotate in other directions so that the CME nose is not directed radially.  The current ForeCAT model cannot account for these rotations, however they can be incorporated in the future through additional parameters describing the CME orientation.  For a more thorough description of ForeCAT see \citet{Kay15}.

To describe the background magnetic field we use the Potential Field Source Surface (PFSS, \citet{Alt69} and \citet{Sch69}), a commonly used model that represents the magnetic field as the gradient of a magnetic potential.  However, based on observations of Type II radio bursts, \citet{Man03} and \citet{Eva08} suggest that the PFSS magnetic field may decay too rapidly with distance.  The standard PFSS model uses a source surface radius, $\Rss$, of 2.5 $\rsun$, beyond which the field is entirely radial and must fall as $R^{-2}$ in the corotating frame.  This source surface distance reproduces the global structure of the solar magnetic field \citep{Hoe84}, but may not accurately describe the ARs at low heights.  We will also explore results using $\Rss$ = 3 $\rsun$, which causes the magnetic field to fall less rapidly with radial distance. 

\section{Results}
We simulate 100 CMEs sampling different masses and final propagation speeds.  The SOHO/LASCO CME catalog \citep{Gop09LASCO} shows a range of CME masses between 10$^{13}$ g and 10$^{16}$ g.  \citet{Vou10} determine an average CME mass of 1.6x10$^{15}$ g from observations of 7668 CMEs.  We simulate CMEs with masses between 10$^{14}$ g and 10$^{16}$ g, ignoring the lowest mass CMEs as they tend to be the least relevant for space weather effects.  For each CME we assume a constant mass with distance, although coronagraph and interplanetary scintillation observations show that a CME mass can increase with distance due to accumulation of the background solar wind material \citep{How07,Vou10}.  The assumption of a constant mass makes these results upper limits on the total deflection and rotation.  LASCO CMEs have speeds between a few tens of km s$^{-1}$ to a few thousand km s$^{-1}$ with an average value of 475 km s$^{-1}$ \citet{Gop09LASCO}.  We consider CME speeds between 300 km s$^{-1}$ and 1500 km s$^{-1}$.  We simulate the CMEs all the way to 1 AU.

All CMEs are initiated from the same active region (AR) in CR 2029 (April-May 2005, declining phase).  The CMEs begin at an initial latitude and longitude of -15.4$\mydeg$ and 17$\mydeg$ and a tilt of 72$\mydeg$ clockwise from the solar equator.  The CMEs have an angular width of 27.6$\mydeg$ and a cross-sectional radius of 0.01 $\rsun$ ($b$ in Figure 2 of \citet{Kay15}).  

\begin{figure}[!hbtp]
\includegraphics[width=6.5in, angle=0]{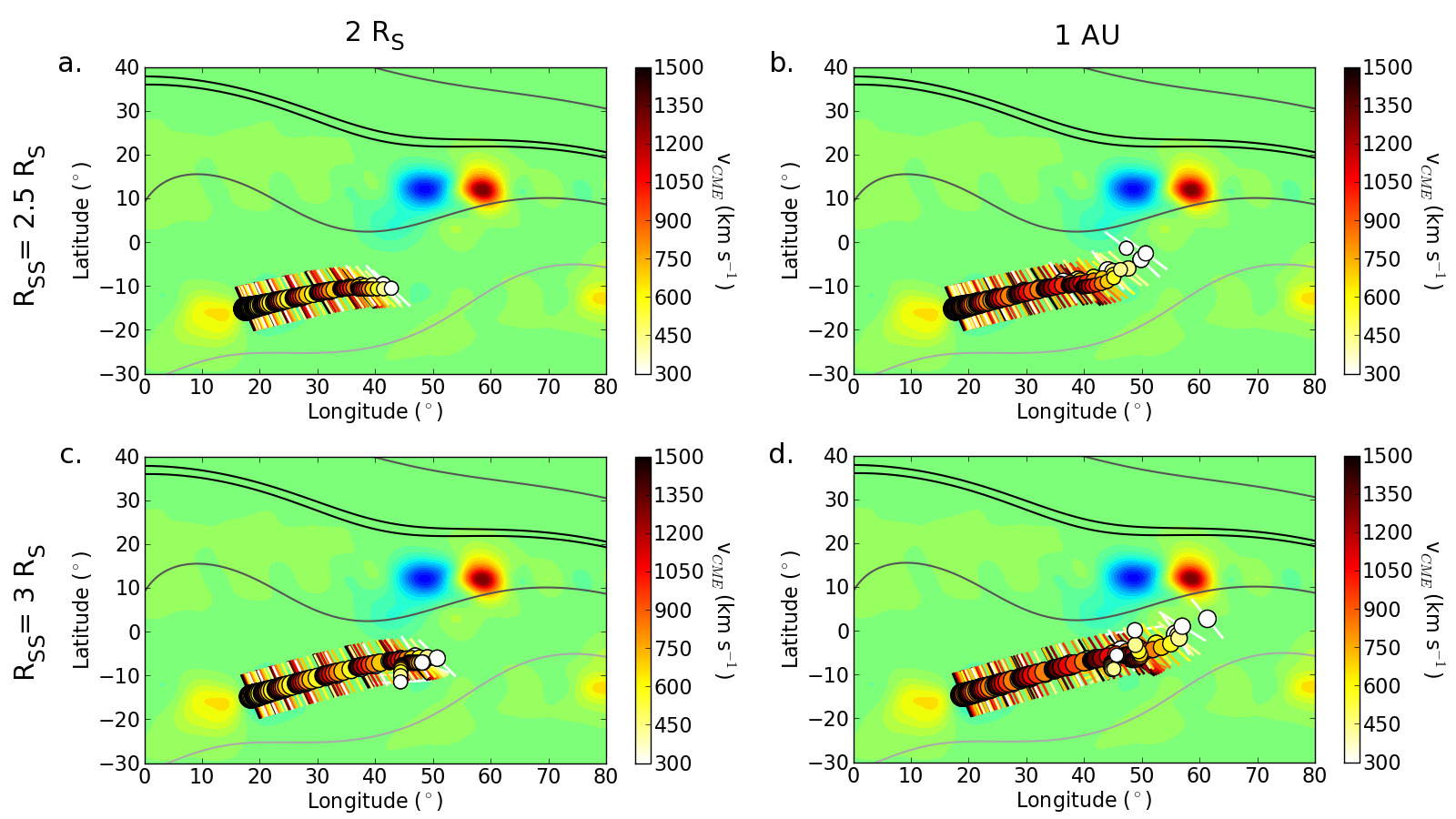}
\caption{Deflected CME position at 2 $\rsun$ (a) and 1 AU (b) for the standard PFSS background 2 $\rsun$ (c) and 1 AU (d) for a source surface radius of 3 $\rsun$.  See description in text. }\label{fig:pos}
\end{figure}

We investigate how close to the Sun a CME's trajectory is determined.  Figure \ref{fig:pos} shows the position of the deflected CMEs at several distances.  The background color contours represent the radial magnetic field at the surface of the Sun.  The large red and blue region near the bottom left of each panel corresponds to the AR from which all the simulated CMEs erupt.  The line contours indicate the radial magnetic field strength farther out.  The CMEs can be deflected initially by imbalances in the AR magnetic field and on global scales the CMEs tend to deflect toward the HCS, the minimum in the background magnetic field intensity (black lines).  Each circle represents the position of an individual CME.  The size of the circle represents the CME mass with larger circles corresponding to more massive ones.  The color of the circle represents the CME's radial speed as indicated by the color bar.  The line through each circle indicates the orientation of each CMEs' toroidal axis.

Figure \ref{fig:pos}(a) and (b) show results using the standard PFSS model with $\Rss$ = 2.5 $\rsun$.  Comparison between the position of the CMEs at 2 $\rsun$ and 1 AU, panels (a) and (b) respectively, shows that the fraction of the deflection beyond 2 $\rsun$ is a small component of the total amount.  The CMEs initially deflect towards the west due to the gradients present in the AR.  As the CMEs propagate outward the gradients from the AR weaken and the global gradients deflect the CMEs northward toward the HCS.  While all CMEs in this case show deflection toward the HCS, none of them reach it.  The slowest, lowest mass CMEs are the most susceptible to variations in the direction of the magnetic gradients.  When the nearby global gradients differ significantly from the local gradients at the initial position they can change the direction of the deflection of the slowest, least massive CMEs.  This behavior corresponds to a rapid increase in the angular momentum, followed by a decrease and finally a gradual increase at farther distances.

As with the deflection, the majority of the rotation occurs below 2 $\rsun$, and the rotation tends to increase with decreasing CME mass and velocity.  The CMEs, initially aligned with the AR polarity inversion line, rotate toward an orientation parallel to the HCS.

Figure \ref{fig:pos}(c) and (d) show results with $\Rss$ = 3 $\rsun$.  The increase in the magnetic field strength causes an increase in the deflection and rotation at both distances.  Despite the increase in the deflection, none of these CME reach the HCS.

We seek to quantify the distance beyond which there would be a negligible change in the CME's position at 1 AU if we do not include the deflection forces beyond this distance.   We know the CME's angular momentum rapidly increases close to the Sun, and the magnetic forces quickly become negligible, after which the CME deflects at a rate corresponding to constant angular momentum.  Similarly, we can describe the continued rotation as the result of angular momentum conservation.

\citet{Kay15} show that when a CME deflects with constant angular momentum its angular position as a function of radial distance, $\theta(\mathrm{R})$, can be described as 
\begin{equation}\label{eq:thetadef}
\theta(\mathrm{R}) = \theta_0 + \frac{\mathrm{v}_{\mathrm{nr,0}}\mathrm{R}_{\mathrm{0}}}{\mathrm{v}_{\mathrm{r}}} \left(\frac{1}{\mathrm{R}_{\mathrm{0}}} - \frac{1}{\mathrm{R}}\right)
\end{equation}
where $\theta_0$ is the angular position of the CME when it begins deflecting with constant angular momentum at a distance $\mathrm{R}_0$ and $\mathrm{v}_{\mathrm{nr,0}}$ and $\mathrm{v}_{\mathrm{r}}$ are the non-radial and radial CME speeds at $\mathrm{R}_0$. The derivation of Equation \ref{eq:thetadef} assumes that the radial speed remains constant.  As r increases $\theta$ asymptotes to a constant value.

\subsection{Deflection and Angular Momentum}
\begin{figure}[!hbtp]
\includegraphics[width=6.5in, angle=0]{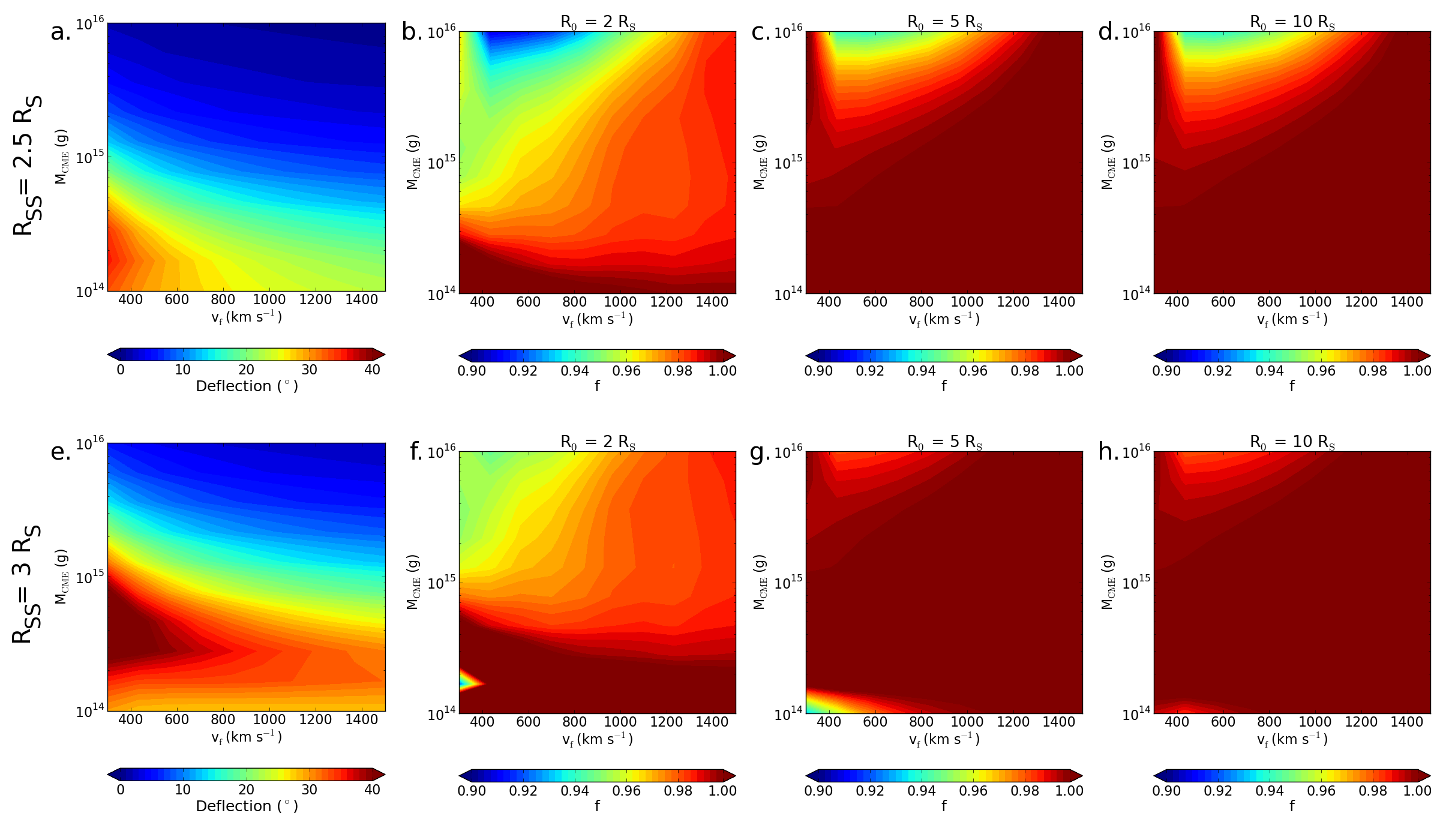}
\caption{The total deflection at 1 AU ((a) and (e)) and the ratio of the predicted to the total deflection for different values of $\mathrm{R}_0$ ((b)-(d) and (f) - (h)).  The top and bottom rows correspond to source surface radii of 2.5 $\rsun$ and 3 $\rsun$.}\label{fig:def}
\end{figure}

We determine how accurately Equation \ref{eq:thetadef} describes the deflection for different values of $\mathrm{R_0}$  For each $\mathrm{R}_0$ we determine the ratio, $f$, of the total deflection at 1 AU predicted by Equation \ref{eq:thetadef}, $\theta(\mathrm{215} \, \rsun)$, and the total simulated deflection at 1 AU, $\theta_{\mathrm{sim}}$, as
\begin{equation}\label{eq:f}
\mathrm{f} = \frac{\theta(\mathrm{215} \, \rsun)}{\theta_{\mathrm{sim}}}
\end{equation} 

Figure \ref{fig:def} shows the total deflection at 1 AU and $f$ for $\mathrm{R_0}$ equal to 2 $\rsun$, 5 $\rsun$, and 10 $\rsun$.  The top row shows results for the standard PFSS magnetic background.  Figure \ref{fig:rot}(a) shows that assuming constant angular momentum beyond 2 $\rsun$ leads to underpredicting the deflection less than 10\% for all masses and velocities.  This underprediction paritally comes from the small continued increase in the angular momentum and the fact the CME's radial speed increases until 3 $\rsun$, but we use the final propagation velocity for Figure \ref{fig:rot}.  This underestimates the deflection between 2 and 3 $\rsun$ as the CME actually propagates slower at this distance.  Figure \ref{fig:def}(c) shows that for $\mathrm{R_0}$ = 5 $\rsun$ the assumption of constant angular momentum yields underpredictions of 1\% to 5\%.  Increasing $\mathrm{R_0}$ to 10 $\rsun$ has little effect.  The largest errors occur for slow, high mass CMEs, which gain little angular momentum in the low corona but slowly gain more at farther distances.  However, these CMEs exhibit deflections of less than 5$\mydeg$ so the the underprediction of 5\% corresponds to only 0.25$\mydeg$.

Figures \ref{fig:def}(e)-(h) show the same as Figures \ref{fig:def}(a)-(d) but for $\Rss$ = 3 $\rsun$.  The increase in the magnetic field strength close to the Sun causes a larger fraction of the angular momentum to be obtained below 2 $\rsun$, causing $f$ to increase for most CMEs at all distances.  The low values of $f$ for small masses occur when a CME's angular momentum decreases as the strength of the global gradients begin to exceed the local gradients and the CME changes direction.  For this source surface radius, the deflection of nearly all the CME can be described within 1\% by assuming constant angular momentum beyond 5 $\rsun$.

\subsection{Rotation and Angular Momentum}
\begin{figure}[!hbtp]
\includegraphics[width=6.5in, angle=0]{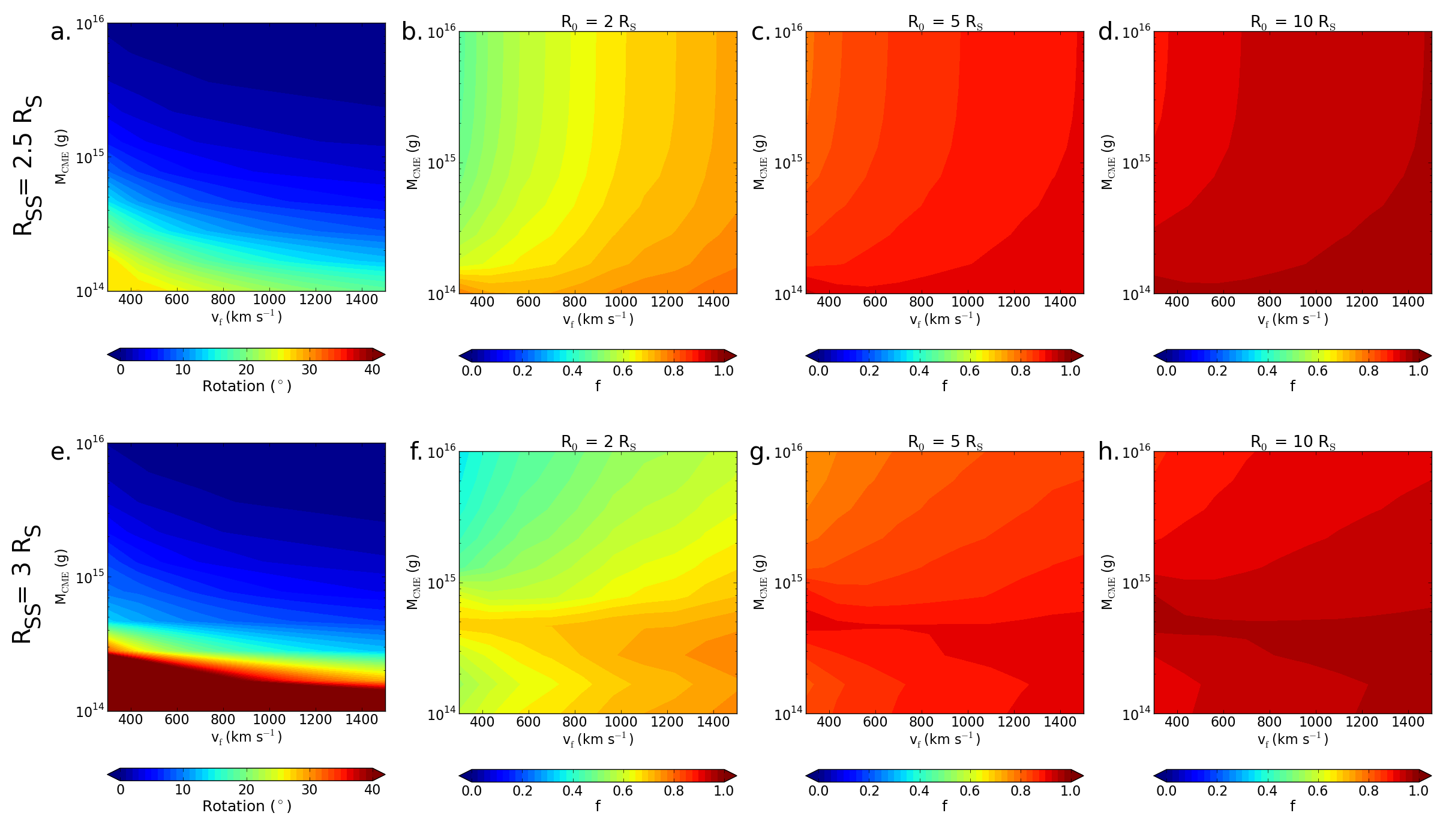}
\caption{Same as Figure \ref{fig:def} but for the rotation.  Note the difference in the contour range.}\label{fig:rot}
\end{figure}

The rotation can also be described by Equation \ref{eq:thetadef} since the moment of inertia for rotation about the CME nose (see \citet{Kay15}) can be shown to be proportional to $\mathrm{R}^2$, assuming self-similar expansion. The non-radial velocity $\mathrm{v}_{\mathrm{nr,0}}$ is replaced by the angular velocity times the distance, $\omega_0\mathrm{R_0}$.  Figure \ref{fig:rot} shows the total rotation and $f$ for several distances for both values of $\Rss$, analogous to Figure \ref{fig:def}. 

The rotational angular momentum tends to noticeably increase out to farther distances than the angular momentum corresponding to deflection.  Only 50\%-80\% of the total rotation is recovered by assuming constant angular momentum beyond 2 $\rsun$.  Assuming constant angular momentum beyond 10 $\rsun$ yields underestimates of the total rotation by 10\%.  The larger source surface causes larger rotations but the behavior with distance is nearly the same for the two source surface heights.  For the slowest, low mass CMEs an error of 10\% may be significant as corresponds to 2.7$\mydeg$ and 11$\mydeg$, for a $\Rss$ of 2.5 $\rsun$ and 3 $\rsun$, respectively.  For the slowest, high mass CMEs this error is negligible as it corresponds to less than 0.1$\mydeg$.

\section{Discussion and Conclusion}
\begin{figure}[!hbtp]
\includegraphics[width=6.5in, angle=0]{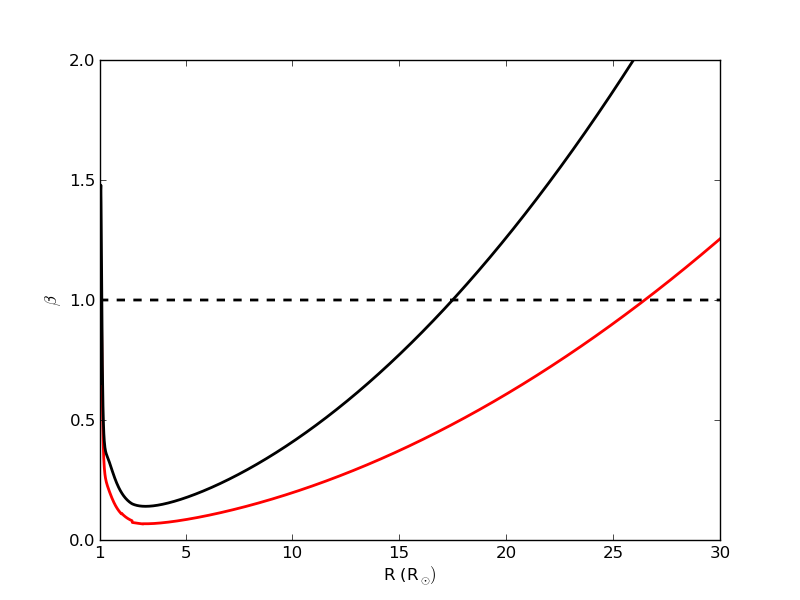}
\caption{Plasma $\beta$ (ratio of thermal to magnetic pressure) versus radial distance for $\Rss$ = 2.5 $\rsun$ (black) or 3 $\rsun$ (red).  The dashed line indicates $\beta$=1.}\label{fig:beta}
\end{figure}

The magnetic forces driving CME deflection and rotation decay rapidly with distance causing little acceleration beyond 2 $\rsun$.  The CME deflects beyond this distance at a rate corresponding to constant angular momentum, asymptotically approaching a constant displacement.  The total simulated deflection at 1 AU can be predicted within 1\% for most CMEs by assuming a CME propagates with constant angular momentum beyond 5 $\rsun$.  The rotation tends to evolve out to farther distances but can be predicted within 10\% by assuming constant angular momentum beyond 10 $\rsun$. We note that these distances are representative of the distance at which the solar wind transitions from a low to a high plasma $\beta$, defined as the ratio of the thermal to magnetic pressure. The solar wind can only efficiently transfer angular momentum to a CME through magnetic forces in a low plasma $\beta$ environment, analogous to the transfer of angular momentum to the solar wind \citep{Web67}.  Figure \ref{fig:beta} shows the plasma $\beta$ versus radial distance above the AR considered in this work.  We use the \citet{Guh06} density model and both versions of the PFSS magnetic field.  ForeCAT does not require a coronal temperature so we assume a constant value of 3 MK, representative of the observed electron temperature above ARs \citep{Ste97}.  Figure \ref{fig:beta} shows that $\beta$ exceeds unity above 17$\rsun$ to 26$\rsun$, with the distance being farther for larger source surface distances.

CME deflection varies according to the relative positions of the HCS, ARs, coronal holes, and CME source region.  The HCS is flat at solar minimum and warped at solar maximum.  Throughout the solar cycle the relative importance of the local and global gradients may change as ARs become more numerous and stronger and the inclination of the HCS increases.  Both factors may affect the distance at which CME deflection is determined; this work has only considered a declining phase Carrington Rotation. 

While many authors have presented observations of interplanetary CME deflections they do not explicitly present the angular momentum at these distances, although it could be estimated from the published trajectories and white-light masses.  If an observed interplanetary deflection has increasing angular momentum, some force must be actively accelerating the CME at interplanetary distances.  Much of observed interplanetary deflection occurs in the longitudinal direction \citep{Lug10, Dav13, Wan14}.  The upcoming Solar Orbiter mission will reach as high as 34$\mydeg$ heliographic latitude and as close as 0.28 AU heliocentric distance, providing an unprecedented view of longitudinal deflections.  This perspective will allow for a more precise study of the evolution of CMEs angular momentum.

ForeCAT includes the magnetic forces at all distances, including interplanetary space.  Our results suggest that the interplanetary magnetic forces are not strong enough to influence the CMEs motion at interplanetary distances with high plasma $\beta$ parameter.  ForeCAT does include many simplifications, notably the lack of enhancement of the solar wind magnetic field surrounding the CME due to the CME's expansion and propagation.  This effect will increase the magnetic deflection forces at interplanetary distances, however, ForeCAT's current interplanetary forces are many orders of magnitude too small to produce noticeable interplanetary deflections.  The magnetic forces at 50 $\rsun$ tend to be about 10$^{-5}$ their coronal values so the compressed magnetic field surrounding the interplanetary CME would need to be enhanced by a factor of over 300 times the ambient value.  We suggest that interplanetary deflections at rates corresponding to increasing angular momentum must be accelerated by non-magnetic forces or result from the interaction of multiple CMEs \citep{Xio06b, Xio09, Lug12}.

The interaction of CMEs with the HCS remains an important area of open research.  The HCS can interfere with the propagation of interplanetary shocks \citep{Ods96}, and will likely also affect CME propagation.  None of the CMEs originating in the AR considered in the work can reach the HCS, however \citet{Kay15} show cases where the CME cross underneath the cusp separating the streamer region from the base of the HCS. Whether the CME can penetrate the HCS farther out may depend on the distance of their interaction.  Proper treatment of this interaction requires a more realistic description of the HCS.  Observations suggest the HCS has a width of 10,000 km (6.7$\times$10$^{-5}$ AU) at 1 AU, with the surrounding plasma sheet being 30 times thicker \citep{Win94,Smi01}.  ForeCAT uses the \citet{Guh06} density model, which creates a 56$\mydeg$ wide density enhancement surrounding the location of the HCS, or a width of 1.48 AU at 1 AU. This scale is comparable to the resolution obtained with MHD models \citep{Oph04}, but ForeCAT does not capture the microphysics at the scale of the actual HCS.  Additionally the impact of the CME upon the HCS will compress and distort the HCS magnetic field producing magnetic forces resisting the CME's passage.  ForeCAT cannot currently capture these effects, but with additional modifications we will address the CME-HCS interaction. 

Since the deflection and rotation tend to be determined by 10 $\rsun$ it is essential to use accurate representations of the solar conditions in this distance range.  Unfortunately this corresponds to the distance at which the current solar models are the most uncertain.  The PFSS magnetic field model, a very commonly used model, assumes that the magnetic field is current-free and can be described as the gradient of a magnetic potential.  The intense magnetic field fields of ARs, which can contribute significantly to the CME deflection, are certainly more complex than this simple current-free approximation.  Additionally, the PFSS model tends to be driven by synoptic maps acquired over a full solar rotation and ARs can evolve on much shorter scales.  These factors also apply to the global magnetic field configuration, but tend to have less of an effect.

Our understanding of the solar magnetic field will greatly improve through the observations by Solar Probe Plus, scheduled to launch in 2018 and reach the smallest perihelion of 8.5 $\rsun$ over six years later.  One of the primary science goals of Solar Probe Plus is to ``determine the structure and dynamics of the magnetic fields at the sources of solar wind.''  Measuring the magnetic field at these close distances will help greatly constrain our magnetic field models.  In the meantime, we suggest that the ForeCAT model can not only reproduce the observed deflection, but also constrain the unknown mass and drag coefficient as well as the background magnetic field.

\acknowledgements  
The authors thank the anonymous referee for the comments.


\begin{thebibliography}{}
\expandafter\ifx\csname natexlab\endcsname\relax\def\natexlab#1{#1}\fi

\bibitem[{{Altschuler} \& {Newkirk}(1969)}]{Alt69}
{Altschuler}, M.~D., \& {Newkirk}, G. 1969, \solphys, 9, 131

\bibitem[{{Byrne} {et~al.}(2010){Byrne}, {Maloney}, {McAteer}, {Refojo}, \&
  {Gallagher}}]{Byr10}
{Byrne}, J.~P., {Maloney}, S.~A., {McAteer}, R.~T.~J., {Refojo}, J.~M., \&
  {Gallagher}, P.~T. 2010, Nature Communications, 1, 74

\bibitem[{{Cremades} \& {Bothmer}(2004)}]{Cre04}
{Cremades}, H., \& {Bothmer}, V. 2004, \aap, 422, 307

\bibitem[{{Davies} {et~al.}(2013){Davies}, {Perry}, {Trines}, {Harrison},
  {Lugaz}, {M{\"o}stl}, {Liu}, \& {Steed}}]{Dav13}
{Davies}, J.~A., {Perry}, C.~H., {Trines}, R.~M.~G.~M., {et~al.} 2013, \apj,
  777, 167

\bibitem[{{Evans} {et~al.}(2008){Evans}, {Opher}, {Manchester}, \&
  {Gombosi}}]{Eva08}
{Evans}, R.~M., {Opher}, M., {Manchester}, IV, W.~B., \& {Gombosi}, T.~I. 2008,
  \apj, 687, 1355

\bibitem[{{Gopalswamy} {et~al.}(2008){Gopalswamy}, {Akiyama}, {Yashiro},
  {Michalek}, \& {Lepping}}]{Gop08}
{Gopalswamy}, N., {Akiyama}, S., {Yashiro}, S., {Michalek}, G., \& {Lepping},
  R.~P. 2008, Journal of Atmospheric and Solar-Terrestrial Physics, 70, 245

\bibitem[{{Gopalswamy} {et~al.}(2009){Gopalswamy}, {Yashiro}, {Michalek},
  {Stenborg}, {Vourlidas}, {Freeland}, \& {Howard}}]{Gop09LASCO}
{Gopalswamy}, N., {Yashiro}, S., {Michalek}, G., {et~al.} 2009, Earth Moon and
  Planets, 104, 295

\bibitem[{{Guhathakurta} {et~al.}(2006){Guhathakurta}, {Sittler}, \&
  {Ofman}}]{Guh06}
{Guhathakurta}, M., {Sittler}, E.~C., \& {Ofman}, L. 2006, Journal of
  Geophysical Research (Space Physics), 111, 11215

\bibitem[{{Gui} {et~al.}(2011){Gui}, {Shen}, {Wang}, {Ye}, {Liu}, {Wang}, \&
  {Zhao}}]{Gui11}
{Gui}, B., {Shen}, C., {Wang}, Y., {et~al.} 2011, \solphys, 271, 111

\bibitem[{{Hildner}(1977)}]{Hil77}
{Hildner}, E. 1977, in Astrophysics and Space Science Library, Vol.~71, Study
  of Travelling Interplanetary Phenomena, ed. M.~A. {Shea}, D.~F. {Smart}, \&
  S.~T. {Wu}, 3--20

\bibitem[{{Hoeksema}(1984)}]{Hoe84}
{Hoeksema}, J.~T. 1984, PhD thesis, Stanford Univ., CA.

\bibitem[{{Howard} {et~al.}(2007){Howard}, {Fry}, {Johnston}, \&
  {Webb}}]{How07}
{Howard}, T.~A., {Fry}, C.~D., {Johnston}, J.~C., \& {Webb}, D.~F. 2007, \apj,
  667, 610

\bibitem[{{Isavnin} {et~al.}(2013){Isavnin}, {Vourlidas}, \& {Kilpua}}]{Isa13}
{Isavnin}, A., {Vourlidas}, A., \& {Kilpua}, E.~K.~J. 2013, \solphys, 284, 203

\bibitem[{{Isavnin} {et~al.}(2014){Isavnin}, {Vourlidas}, \& {Kilpua}}]{Isa14}
---. 2014, \solphys, 289, 2141

\bibitem[{{Kay} {et~al.}(2015{\natexlab{a}}){Kay}, {dos Santos}, \&
  {Opher}}]{Kay15L}
{Kay}, C., {dos Santos}, L.~F.~G., \& {Opher}, M. 2015{\natexlab{a}}, \apjl,
  801, L21

\bibitem[{{Kay} {et~al.}(2013){Kay}, {Opher}, \& {Evans}}]{Kay13}
{Kay}, C., {Opher}, M., \& {Evans}, R.~M. 2013, \apj, 775, 5

\bibitem[{{Kay} {et~al.}(2015{\natexlab{b}}){Kay}, {Opher}, \& {Evans}}]{Kay15}
---. 2015{\natexlab{b}}, ArXiv e-prints, arXiv:1410.4496

\bibitem[{{Kilpua} {et~al.}(2009){Kilpua}, {Pomoell}, {Vourlidas}, {Vainio},
  {Luhmann}, {Li}, {Schroeder}, {Galvin}, \& {Simunac}}]{Kil09}
{Kilpua}, E.~K.~J., {Pomoell}, J., {Vourlidas}, A., {et~al.} 2009, Annales
  Geophysicae, 27, 4491

\bibitem[{{Liu} {et~al.}(2010){Liu}, {Thernisien}, {Luhmann}, {Vourlidas},
  {Davies}, {Lin}, \& {Bale}}]{Liu10b}
{Liu}, Y., {Thernisien}, A., {Luhmann}, J.~G., {et~al.} 2010, \apj, 722, 1762

\bibitem[{{Lugaz}(2010)}]{Lug10a}
{Lugaz}, N. 2010, \solphys, 267, 411

\bibitem[{{Lugaz} {et~al.}(2012){Lugaz}, {Farrugia}, {Davies}, {M{\"o}stl},
  {Davis}, {Roussev}, \& {Temmer}}]{Lug12}
{Lugaz}, N., {Farrugia}, C.~J., {Davies}, J.~A., {et~al.} 2012, \apj, 759, 68

\bibitem[{{Lugaz} {et~al.}(2010){Lugaz}, {Hernandez-Charpak}, {Roussev},
  {Davis}, {Vourlidas}, \& {Davies}}]{Lug10}
{Lugaz}, N., {Hernandez-Charpak}, J.~N., {Roussev}, I.~I., {et~al.} 2010, \apj,
  715, 493

\bibitem[{{MacQueen} {et~al.}(1986){MacQueen}, {Hundhausen}, \&
  {Conover}}]{Mac86}
{MacQueen}, R.~M., {Hundhausen}, A.~J., \& {Conover}, C.~W. 1986, \jgr, 91, 31

\bibitem[{{Mann} {et~al.}(2003){Mann}, {Klassen}, {Aurass}, \&
  {Classen}}]{Man03}
{Mann}, G., {Klassen}, A., {Aurass}, H., \& {Classen}, H.-T. 2003, \aap, 400,
  329

\bibitem[{{Nieves-Chinchilla} {et~al.}(2013){Nieves-Chinchilla}, {Vourlidas},
  {Stenborg}, {Savani}, {Koval}, {Szabo}, \& {Jian}}]{Nie13}
{Nieves-Chinchilla}, T., {Vourlidas}, A., {Stenborg}, G., {et~al.} 2013, \apj,
  779, 55

\bibitem[{{Odstr{\v c}il} {et~al.}(1996){Odstr{\v c}il}, {Dryer}, \&
  {Smith}}]{Ods96}
{Odstr{\v c}il}, D., {Dryer}, M., \& {Smith}, Z. 1996, \jgr, 101, 19973

\bibitem[{{Opher} {et~al.}(2004){Opher}, {Liewer}, {Velli}, {Bettarini},
  {Gombosi}, {Manchester}, {DeZeeuw}, {Toth}, \& {Sokolov}}]{Oph04}
{Opher}, M., {Liewer}, P.~C., {Velli}, M., {et~al.} 2004, \apj, 611, 575

\bibitem[{{Schatten} {et~al.}(1969){Schatten}, {Wilcox}, \& {Ness}}]{Sch69}
{Schatten}, K.~H., {Wilcox}, J.~M., \& {Ness}, N.~F. 1969, \solphys, 6, 442

\bibitem[{{Shen} {et~al.}(2011){Shen}, {Wang}, {Gui}, {Ye}, \& {Wang}}]{She11}
{Shen}, C., {Wang}, Y., {Gui}, B., {Ye}, P., \& {Wang}, S. 2011, \solphys, 269,
  389

\bibitem[{{Smith}(2001)}]{Smi01}
{Smith}, E.~J. 2001, \jgr, 106, 15819

\bibitem[{{Sterling} {et~al.}(1997){Sterling}, {Hudson}, \& {Watanabe}}]{Ste97}
{Sterling}, A.~C., {Hudson}, H.~S., \& {Watanabe}, T. 1997, \apjl, 479, L149

\bibitem[{{Vourlidas} {et~al.}(2011){Vourlidas}, {Colaninno},
  {Nieves-Chinchilla}, \& {Stenborg}}]{Vou11}
{Vourlidas}, A., {Colaninno}, R., {Nieves-Chinchilla}, T., \& {Stenborg}, G.
  2011, \apjl, 733, L23

\bibitem[{{Vourlidas} {et~al.}(2010){Vourlidas}, {Howard}, {Esfandiari},
  {Patsourakos}, {Yashiro}, \& {Michalek}}]{Vou10}
{Vourlidas}, A., {Howard}, R.~A., {Esfandiari}, E., {et~al.} 2010, \apj, 722,
  1522

\bibitem[{{Wang} {et~al.}(2004){Wang}, {Shen}, {Wang}, \& {Ye}}]{Wan04}
{Wang}, Y., {Shen}, C., {Wang}, S., \& {Ye}, P. 2004, \solphys, 222, 329

\bibitem[{{Wang} {et~al.}(2014){Wang}, {Wang}, {Shen}, {Shen}, \&
  {Lugaz}}]{Wan14}
{Wang}, Y., {Wang}, B., {Shen}, C., {Shen}, F., \& {Lugaz}, N. 2014, Journal of
  Geophysical Research (Space Physics), 119, 5117

\bibitem[{{Weber} \& {Davis}(1967)}]{Web67}
{Weber}, E.~J., \& {Davis}, Jr., L. 1967, \apj, 148, 217

\bibitem[{{Winterhalter} {et~al.}(1994){Winterhalter}, {Smith}, {Burton},
  {Murphy}, \& {McComas}}]{Win94}
{Winterhalter}, D., {Smith}, E.~J., {Burton}, M.~E., {Murphy}, N., \&
  {McComas}, D.~J. 1994, \jgr, 99, 6667

\bibitem[{{Xiong} {et~al.}(2009){Xiong}, {Zheng}, \& {Wang}}]{Xio09}
{Xiong}, M., {Zheng}, H., \& {Wang}, S. 2009, Journal of Geophysical Research
  (Space Physics), 114, 11101

\bibitem[{{Xiong} {et~al.}(2006{\natexlab{a}}){Xiong}, {Zheng}, {Wang}, \&
  {Wang}}]{Xio06a}
{Xiong}, M., {Zheng}, H., {Wang}, Y., \& {Wang}, S. 2006{\natexlab{a}}, Journal
  of Geophysical Research (Space Physics), 111, 8105

\bibitem[{{Xiong} {et~al.}(2006{\natexlab{b}}){Xiong}, {Zheng}, {Wang}, \&
  {Wang}}]{Xio06b}
---. 2006{\natexlab{b}}, Journal of Geophysical Research (Space Physics), 111,
  11102

\bibitem[{{Xiong} {et~al.}(2007){Xiong}, {Zheng}, {Wu}, {Wang}, \&
  {Wang}}]{Xio07}
{Xiong}, M., {Zheng}, H., {Wu}, S.~T., {Wang}, Y., \& {Wang}, S. 2007, Journal
  of Geophysical Research (Space Physics), 112, 11103

\bibitem[{{Zhang} \& {Dere}(2006)}]{Zha06}
{Zhang}, J., \& {Dere}, K.~P. 2006, \apj, 649, 1100

\end{thebibliography}

\end{document}